# Decoding Financial Behaviour: An Analysis of urbanised households in India using AIDIS 77[th] round.

**Dr Divya Sharma**
**Assistant Professor**
**VIPS-TC, GGSIPU, Delhi**
**&**
**PhD, Ambedkar University Delhi**
divya.sharma@vips.edu
Ph. No-9717759366

## Abstract:

This research paper delves into the financial behavior of urbanized households in India, specifically focusing on million-plus agglomerations. Using data from the 77th round of the All India Debt and Investment Survey, the study analyzes assets, borrowing patterns, and the usage of financial instruments like bank accounts, e-wallets, and life insurance. The research explores demographic factors, household structures, and district-level parameters to understand the intricate financial landscape. With a focus on low-income households, the study identifies income, education, type of employment, and branches per capita as significant factors influencing financial behavior. The study contributes valuable insights for financial institutions, policymakers, and researchers seeking a comprehensive understanding of the financial lives of urban households in India.

## Introduction:

This paper endeavors to address a crucial gap in our understanding of financial dynamics of most urbanized Indian households that is the million plus agglomerations. It focuses on million-plus agglomerations which acknowledges the presence of financial services, prompting a need to scrutinize the demand-side factors influencing usage patterns. The paper aims to unravel the intricate financial landscape of urbanized Indian households through a meticulous examination of data sourced from the latest 77th round of the All India Debt and Investment Survey, coupled with a comparative analysis against the preceding 70th round. Our focus is on understanding assets and borrowing patterns.

The research explores factors influencing the usage of bank accounts and e-wallets. It scrutinizes asset distribution among saving accounts, gold, bonds, and LICs. It explores the factors influencing the usage of bank accounts and e-wallets. Additionally, we delve into the distribution of assets among saving accounts, gold, bonds, and LICs. The study further categorizes loans based on major borrowing purposes and investigates the factors affecting housing loans, business loans, and other expenditure loans. This multifaceted analysis not only strives to comprehend the intricate fabric of urban financial behavior but also contributes valuable information for financial institutions, policymakers, and researchers seeking a comprehensive understanding of the financial lives of urban households in India.

The paper looks at the following: Firstly, we carefully document the assets and debt composition of the most urban Indian households that are among the million plus agglomerations in the economy. Secondly, we look at both the usage and intensity of usage of the most important financial instruments used by households, namely bank accounts, e-wallets and life insurance and others. Thirdly, we highlight the significant features of the debt of urban Indian households as well as evaluate the demographic factors which affect households taking loans or not.

Then we evaluate the major purposes for which loan is taken, whether the source of loan is institutional or non-institutional. We also try to understand the factors which affect the household in getting an institutional loan for different purposes like housing or business needs.

Talking about the demographic factors we have taken age, education, employment, religion and caste of the household. Then among the household structure, the size of the household and gender of the household head is taken. Here, apart from the demographic characteristics and household structure the residual variations are captured through district level parameters like branches per capita, average literacy and sex-ratio.

We have restricted ourselves to Indian urban agglomerations with a population of more than a million according to the 2011 census. Urban agglomerations refer to contiguous urban areas which includes the city as well its outgrowth. There are fifty-three such agglomerations according to the 2011 census. We are interested in looking at specific urban areas in the Indian economy where the financial access is highest, so that we can look at the factors which affect the household's finances despite financial access. Then these urban agglomerations have socio-economic conditions which are different from the other urban areas so it made sense to study them separately

We use individual- and household-level data from the most recently released All India Debt and Investment Survey (AIDIS) dataset of the National Sample Survey Office (NSSO) of India, the seventy-seventh round, and the RBI's time series publication in which bank branch statistic is used to understand the significant features of Indian household finance and the determinants of access and usage of the prominent financial instruments. The survey period for the seventy seventh round was from July 2018 to December 2018 for the first visit and the second visit had a time frame from January 2019 till June 2019.

The All India Debt and Investment Survey is one of the oldest decennial statistical exercises conducted by National Sample Survey Office (NSSO) to obtain reliable quantitative information regarding stock of assets, debt, borrowing and capital formation to name a few indicators of the Indian households. For measuring the availability of formal financial services we make use of the RBI's bank branch statistics which provides region/state and district wise data regarding functional bank branches in the economy. The data is released annually. We take data from 2018-19 to ensure comparability with the AIDIS dataset. Data on literacy rates and sex ratios are taken from the Census 2011.

Since the aim of this thesis is to study the financial lives of low-income households, we have removed the data for the households belonging to the top most consumption quartile from the AIDIS data we use.

We need to understand the demographic factors which affect the saving and credit behaviour of the households. The probable demographic factors on which these variations can depend are education, family composition, social groupings in the form of caste, religion and others, along with income and wealth held. Even after including all the household demographic characteristics there is scope for considerable residual variation which can be addressed by looking at the district or state variations in variables like bank branches per capita, average literacy, sex ratio and other socio-economic indicators of the district.

Now this study finds out that income, education, type of employment and branches per capita are the main significant factors which affect the financial behaviour of the low-income households living in the million plus urban agglomerations. We find a positive association of income and education with savings accounts but a negative relationship with e-wallets usage.

Since we have bifurcated the loans into three main categories that is housing, business and other household expenditures, we find that housing and business loans proportions increase with income and education but we find a decline in the proportion of households expenditure loans with rise in income and education levels. Same holds true for the type of employment categories. So we see that same three factors are significant but as the purpose of loan changes then the effect of the same factors changes

We also find that branches per capita have a negative relationship with bank's saving account and life insurance policies but have a positive relationship in having higher gold deposits. Then from the loans side we find that it has a positive relationship in inducing institutional business loans.

**Literature Review:**

There exists only limited research on the financial lives of Indian households. Most of the literature that does exist deals with the rural sector and there is a dearth of serious study of financial lives of the poor in the urban sector. It is this research gap that has motivated the present study.

Badarinza et. al (2016) uses the AIDIS 70[th] round dataset to look at the significant features of Indian households' financial decisions and compares their asset allocations to those of households in other countries using international micro-data. The study finds that household characteristics like age, education, wealth and rural-urban areas significantly affect the structure of a household's balance sheet. After controlling for household demographics, there are considerable state level variations observed in the assets and debt holdings, which is tried to be explained through state level factors like share of population in the public sector, bank branch networks and historical inflation volatility.

Then density of bank branches across the Indian states is used as a supply side indicator and it is found out that households in states which have high bank branch penetration rely less on non-institutional debt. It is also found out that education plays an important role in shifting a household's asset allocation towards financial assets and is also associated with a shift towards low cost institutional debt sources from the high dependence on non-institutional debt in general (Badarinza et. al. 2016)

RBI's Committee on Household Finance (RBI 2017) in its report made use of multiple rounds of All India Debt and Investment Survey (AIDIS) in India as well as micro data from other countries like Australia, UK, Germany, US, China and Thailand.
The report shows that as wealth of the household increases it leads to further accumulation of physical assets only, but then more of real estate is preferred over gold deposits. It is also found out that even for the wealthiest Indian household the financial asset holdings are modest. This is because physical assets are preferred over financial assets not because of cultural or behavioural factors but because physical assets like gold deposits and real estate are sources for avoiding scrutiny in investment, which is present while investing in other financial assets in the formal sector. Then households with higher income also find it easier to do tax evasion with physical assets as well as a relatively safer place for investing their illicit earnings (RBI. 2017)

The report argues that higher levels of education is associated with a lower share of gold deposits and real estate and an increase in the share of financial assets. This happens because higher education leads to better employment opportunities like the formal sector employment which reduces tax evasion and also education increases financial knowledge as well as awareness to use financial products.

Kumar et. al. (2015) used data from three consecutive rounds of AIDIS, that is $48^{th}$ round (1992), $59^{th}$ round (2003) and $70^{th}$ round (2013), to study changes in the structure of rural credit across social groups and states, as well as the factors affecting the choice between formal and informal sources of credit.
They found that access to credit in rural areas is affected by socio-economic characteristics, education, gender and caste. Their study also found significant regional disparities. The study observes that family size has a negative and significant role, that is, increase in family size reduces the access to formal credit. The age of the household head has a positive and significant effect on the usage of formal financial institutions as with increase in age comes better experience and decision making capacity. There is a gender bias with male headed households finding it more difficult than the female headed households to access credit. Caste also matters, as compared to the general category the scheduled tribe and other backward classes find it difficult to access credit. Thus, overall it is observed that education, caste, age and gender have been found to influence the access of formal financial institutions among rural households (Kumar et. al.2015)

Pradhan (2013) looked at almost all the AIDIS rounds till $70^{th}$ round (2013) and also referred to three reports that is Report of the Technical group to Review legislation on Money Lending (2006), Report of the Taskforce on Credit related issues of Farmers (2010) and the Malegam Committee Report (2011). It concludes that despite several efforts to foster formal credit, though there has been some decline in the dependence of non-institutional credit, the majority of loans still remained as informal finance. The study finds that around two-fifths of the rural households are still dependent on informal financial sources. The paper sees in this an immense scope for financial inclusion in those areas.

Demirgüç-Kunt and Klapper (2013) have examined household finance by looking at the saving and borrowing decision of the households, using the Global Financial Inclusion (Global Findex ) 2011 database. It looked at three things. Firstly, ownership and use of accounts in a formal institution. Secondly looking at the saving behaviour which includes saving at all formal institutions as well as community led saving schemes. Third set of indicators were based on borrowing which included purposes and source of borrowing. It found that saving in formal financial institutions depends on household characteristics.

Using the multivariate probit model Demirgüç-Kunt and Klapper observed that keeping all the other characteristics like household size, employment, gender and education constant, as income increases the usage of formal saving mechanisms increases. This increase is even higher among high income and developed countries as compared to the low income countries. Coming to the borrowings it found that in developed and high income countries borrowing from formal sources is much more than in developing and middle income economies. Among borrowings it found that in the case of new formal loans in the past one year there is statistically significant difference of about 4 percent between the richest and poorest income quartile among developing countries and there is no significant difference among the developed economies (Kunt and Klapper. 2013)

Finance and financial services are the backbone of any economy and the reach of the financial sector is an important indicator of a country's economic development. Financial inclusion has been defined as "the process of ensuring access to financial services, timely and adequate credit for vulnerable groups such as weaker sections and low-income groups at an affordable cost" (Committee on Financial Inclusion- Rangarajan, RBI, 2008). However due to many constraints many households are not able to avail these financial services and are financially excluded. Financial exclusion is defined as "the lack of access by certain segments of the society to appropriate, low cost, fair and safe financial products and services from mainstream providers" by former RBI governor Rakesh Mohan (Mohan, 2006)

Being financially excluded impacts the normal financial lives of the household and they can fall prey to poverty, unemployment and over-indebtedness (Anand & Chhikara, 2013). The study used a log linear model and regressed the determinants and major indicators of financial inclusion across Indian states. It finds out that a one percent increase in financial inclusion leads to 0.139 percent increase in the human development index (Anand & Chhikara, 2013). It highlights that availability and usage of financial services are inversely related with poverty levels.

Kunt and Klapper (2013) have tried to find out the barriers to financial inclusion and it is observed that globally the most cited reason for not using formal accounts is low income or lack of enough money to save, the next reason cited is that opening bank accounts are expensive so even if one member in the household has it then it is enough. Thirdly, among the developing economies, distance, time and cost involved in going to a formal institution is high.

These barriers highlight that the major critique of the financial inclusion strategies is that there is no connect between the financial instruments offered and the prevalent socio-cultural contexts. Mobile money services are a step towards a diverse ecosystem of mobile technology based services (Villasenor. 2013; Gabor and Brooks. 2017)

Growing mobile phone ownership among the low-income economies has created immense opportunities for mobile money and digital payments. To deepen financial access to formal institutions, innovative mechanisms like mobile money and branchless banking are a step towards serving the unbanked and the financially excluded but also to those who underutilize or do not use their bank accounts. Mobile money helps to facilitate formal financial transactions through a mobile phone in a relatively economical, reliable and secure manner (Jack and Suri. 2011; Duncombe and Boateng. 2009)

McKay and Pickens (2010) have done an international comparison between mobile money and traditional banking and found branchless banking to be 19 percent more cheaper. . Kenya's M-pesa was roughly one third to one half as expensive as the other prevailing financial instruments. This lower starting and operational cost of mobile money becomes an added advantage for the poor and vulnerable groups.

Well supervised mobile money is much safer than keeping cash at home specially for the unbanked or those who find it difficult to operate the bank account. Morawczynski (2009) found that the scope for privacy and autonomy is higher in mobile money, especially for vulnerable groups. The study found that access to mobile money has increased women's personal savings as she no longer needed to ask for her husband's permission.
There are advantages in terms of liquidity as well as speed also because in emergencies mobile money which is quite quick comes to the rescue. (Donovan. 2012)

There has been a growing adult population who owns a mobile phone in the Indian economy. Now with the government push towards supply led measures to augment financial inclusion in the form of universal bank account ownership in 2014 as well as promoting cashless payments after the strategic 2016's demonetization exercise also gave a boost to digital payments in the economy (Chandrashekhar and Ghosh. 2018). Mobile money has the potential to transform the financial system as it is most promising among the developing countries. The most successful ones have been M-pesa in Kenya, Gcash in Philippines, overall the broadest success have been seen in the Sub Saharan African region (Kunt and Klapper. 2013). So, proliferation of mobile banking can be a mechanism for enhancing financial access among the vulnerable and underserved groups in the developing economies.

Joshi et. al. (2019) study how digital payment mechanisms can create financial inclusion. It specifically looked at the Paytm wallet awareness and usage among street vendors in urban Indian markets. The study was based on interviews and ethnographic observations amongst the street vendors of five markets in Delhi NCR and in Greater Hyderabad. The aim was to understand their business transactions and evaluate the acceptance of Paytm wallet among the vendors as well as uncover the socio-economic factors behind the patterns of usage. The respondents were in the age group of 22-55 years and were street vendors, micro business owners and waiters. The study found that business type and social framework affects the usage of digital wallets. It also found that financial literacy is an acquired habit borne by everyday usage and there is an enormous role of local actors and social system in it.

**Descriptive Statistics:**

Our primary data source is the All India Debt and Investment survey conducted by the National Statistical Office (NSO) (formerly the National Sample Survey Office ). The All India Debt and Investment Survey is one of the oldest decennial statistical exercises conducted by the NSO/NSSO. It collects data regarding stock of assets, debt, borrowing, capital formation and other indicators from Indian households. We look at the latest 77th round of the survey which was conducted during January to December 2019. The survey was conducted in two visits, first one between January to August 2019 and second round for the same respondents between September to December 2019. The survey covered 69,455 households in 5,940 villages and 47,006 households in 3,995 urban blocks.

RBI's time series dataset, in particular the bank branch statistics which provides region/state and district wise data regarding functional bank branches. This is an annual publication. For compatibility with AIDIS we use data from 2018-19. Data for district-wises literacy rates and sex ratio indicators are taken from Census 2011

Now from these data sets, we pick up the relevant variables for the study. Since the data set does not provide income of the household, we use consumption as a proxy for it. We take the sum of a household's monthly income and then calculate annual income. We then take a log of it and use it as an independent variable in several regression.

Here we compare the two rounds of NSSO's AIDIS data set for the latest seventy-seventh round (2018-19) and the previous one that is seventh round (2013-14).

**Table:1 Comparison of key indicators between round 2018-19 and 2013-14**

| Indicators | Total Urban 2018-19 | Total Urban 2013-14 | Million plus agglomerations 2018-19 |
|---|---|---|---|
| **Bank account holders** | 93.93 | 82.74 | 93.67 |
| **LIC** | 20.76 | 18.5 | 22.55 |
| **E-wallets** | NA | 13.38 | 17.87 |
| **Composition of loans by value of loan** | | | |
| Business loan | 15.79 | 22.62 | 11.47 |
| Medical loan | 3.13 | 2.7 | 2.1 |
| Housing loan | 63.97 | 56.47 | 72.78 |
| Other loan | 17.11 | 18.21 | 13.62 |
| **Composition of loans by number of loans** | | | |
| Business loan | 17.64 | 18.04 | 14.14 |
| Medical loan | 9.89 | 11.02 | 7.87 |
| Housing loan | 26.69 | 25.92 | 31.12 |
| Other loan | 45.78 | 45.03 | 46.86 |

Here we compare the key indicators among the selected million plus agglomerations vis a vis the total urban population in the Indian economy. These values above are the percentages of households using the asset or taking loan. We find that the bank account ownership is almost similar in both places. If we look at the percentage of Life insurance policies (LICs) taken up then we find just a two percent higher insurance take up rate in the urban agglomerations as compared to total urban areas in 2018-19. Looking at the e-wallets we do not have data for 2013-14 as it was not recorded in the AIDIS, but if we compare the total urban with the urban agglomerations then the percentage of e-wallets are around three percent higher in the urban agglomerations.

However some difference is seen among the categories of loan in terms of volume as well as number of loans. The volume of housing loans is much higher in million plus agglomerations. Whereas, the volume of other loans is slightly lower in urban agglomerations as compared to the average urban areas.

If we look at the percentage of households having a bank account in the total urban India, we find that there is roughly close to nine percent increase from 2013 till 2018.
We find that among the volume of loan category, it is the housing loan which holds the maximum percentage in both the years. Among the number of loans disbursed then other loans hold the maximum share in both the respective years.

**Table 2: Major Statistics for the 2018-19 round**

| Indicators | Total Urban | Million plus Agglomerations |
|---|---|---|

| Financial Savings | | |
|---|---:|---:|
| Saving Deposits | 31.90 | 28.47 |
| Fixed Deposits | 11.63 | 14.93 |
| Post Office Deposits | 2.69 | 2.11 |
| Gold and bullions | 52.93 | 53.70 |
| Other financial savings | 0.86 | 0.77 |
| | | |

Next we contrast some of the statistics for total urban India with that of the million plus agglomerations. We look at the types of financial assets owned by the urban Indian households in 2018-19. We find that major ownership lies in holding gold deposits in the form of jewellery or coins as close to fifty three percent of the portfolio is there for the urban Indian households as well as for the million plus agglomerations.

Now among the data sets under consideration, they do not provide income of the household so we use consumption as a proxy for it. We take the sum of a household's monthly income and then calculate annual income. We then take a log of it.

**Table 3: Proportions according to income quartiles**

| Income Quartiles | E-wallets | Bank account |
|---|---|---|
| 1 | 0.89 (0.01) | 0.89 (0.004) |
| 2 | 0.88 (0.01) | 0.93 (0.004) |
| 3 | 0.82 (0.01) | 0.95 (0.003) |
| 4 | 0.73 (0.01) | 0.97 (0.001) |

**Table4: Proportions according to Household type**

| Household type | E-wallet | Bank Account |
|---|---|---|
| Self-employed | 0.84 (0.01) | 0.94 (0.003) |
| Salaried | 0.79 (0.01) | 0.95 (0.002) |
| Casual | 0.95 (0.00) | 0.87 (0.006) |
| Others | 0.84 (0.00) | 0.01 (.003) |

**Table 5: Proportions according to Education**

| Education | E-wallet | Bank account |
|---|---|---|
| Illiterate | 0.98 (0.00) | 0.82 (.009) |
| Primary | 0.96 (0.00) | 0.91 (.003) |
| Secondary | 0.85 (0.01) | 0.96 (.002) |
| Diploma | 0.67 (0.02) | 0.99 (.002) |
| Graduation & above | 0.58 (0.01) | 0.99 (.001) |

**Table 6: Proportions of financial assets according to income quartiles**

| Income Quartiles | Saving Account | Current Account | Fixed Deposit | Post Office | Cooperatives | Pension Fund | Other financial assets | Gold & Bullions |
|---|---|---|---|---|---|---|---|---|
| 1 | 0.21 | 0.00 | 0.03 | 0.02 | 0.02 | 0.05 | 0.00 | 0.67 |
| 2 | 0.20 | 0.00 | 0.03 | 0.01 | 0.01 | 0.08 | 0.01 | 0.66 |
| 3 | 0.21 | 0.01 | 0.03 | 0.02 | 0.01 | 0.09 | 0.01 | 0.64 |
| 4 | 0.24 | 0.01 | 0.04 | 0.02 | 0.01 | 0.13 | 0.00 | 0.56 |
| Total | 0.22 | 0.01 | 0.03 | 0.02 | 0.01 | 0.09 | 0.00 | 0.63 |

From table 3 to 6 we look at the mean estimations of e-wallets, bank accounts and other financial assets on the basis of income quartiles, employment types (household type) and education.

As we know that income plays an important role in the usage of financial instruments. Looking at the asset side first, we find that as income increases some slight improvement in the mean usage of bank accounts increases, however the intensity of usage (proportion of saving account to total financial wealth) remains almost constant around mean value of 0.20 for all income quartiles.

If we look at e-wallets then we find that for the first three quartiles the usage of e-wallet is constant around 0.80 mean value, it slightly falls at the top most quartile, indicating a reduction in e-wallets usage at the top most quartile. A similar pattern is observed in the proportion of gold and bullion with increase in income; a marginal drop is observed in the top most quartile. This means at the top most income level the household must be saving in other instruments apart from the basic ones like gold and bullion, saving account or e-wallets. They must be investing in shares and mutual funds, but studying more about the top most income bracket households isn't in the scope of the study.

Coming to the mean usage among different employment types we find that apart from others all the self-employed and salaried use bank accounts in the same manner with just a marginal less usage among casuals. The pattern is totally opposite when it comes to e-wallet usage as casuals and others use it the most and somewhat lesser is being used by self-employed and salaried. This means that bank accounts and e-wallets can be considered as substitutes.

Education is seen to increase the usage of bank account as the mean usage by illiterates is slightly less than the rest and higher education increases usage, however the pattern is opposite in case of e-wallets because illiterate mean usage is maximum and most educated ones that is graduates and above are using it the least. More education is reducing usage of e-wallets, this could be possible because the year we are discussing here is 2018-19 when the e-wallets were at very primary stages and people were using it for transactional needs only and there was no linkage with bank accounts also. So the more educated ones might not be using it that frequently or not storing money in their wallets rather saving it in other financial instruments. Overall we find that bank accounts and e-wallets can be considered as substitutes.

**Table 7: Proportions of loans taken according to Income Quartiles**

| Income Quartiles | Proportions of total loans | Proportion of total housing | Proportion of total Business | Proportion of total Other household expenditure |
|---|---|---|---|---|
| 1 | 0.91 | 0.18 | 0.15 | 0.55 |
| 2 | 0.90 | 0.27 | 0.15 | 0.49 |
| 3 | 0.87 | 0.33 | 0.14 | 0.47 |
| 4 | 0.87 | 0.43 | 0.13 | 0.39 |

Table 8: **Proportions of Institutional loan taken according to Income Quartiles**

| Income Quartiles | Proportion of Institutional loan | Proportion of Institutional Housing | Proportion of Institutional Business | Proportion of Institutional Other household expenditure |
|---|---|---|---|---|
| 1 | 0.53 | 0.77 | 0.82 | 0.48 |
| 2 | 0.67 | 0.91 | 0.86 | 0.6 |
| 3 | 0.71 | 0.95 | 0.86 | 0.66 |
| 4 | 0.81 | 0.97 | 0.89 | 0.74 |

Now coming to loans, we compare and contrast between the loans in general and institutional loans taken by the households on factors like income, employment and education. Then we also bifurcate and introspect the three main purposes of loan that is housing, business and other household expenditure.

If we analyse how a household takes a loan based on their income quartile. We find that the total proportion of loans falls as income increases and we see the maximum proportion of loans are taken by the lowest income quartile. This reflects that the financial lives of the poor involve more loans, though the amount involved would be lower than their richer counterparts but they keep borrowing to meet their financial needs. However, if we contrast it with the total institutional loans taken up then the trend reverses because the lowest quartile borrows the least from an institutional source. As income increases then institutional borrowing increases too.

If we now move on to the different purposes of loan, we find that the proportion of housing loans increases with increase in income, however business and other household expenditure loans are higher for lower income quartiles indicating a lower need to borrow for household expenditures as income rises. However, when it comes to institutional loans then for each purpose of loan there is a positive relationship with income levels. The richest or top quartile borrows the most from them. So institutional loan sources cater more to the richer than to the poorer ones even among the million plus agglomerations.

**Table 9: Proportion of loans taken according to Household type**

| Household Type | Proportions of total loans | Proportion of total housing | Proportion of total Business | Proportion of total Other household expenditure |
|---|---|---|---|---|
| Self-employed | 0.89 | 0.24 | 0.30 | 0.35 |
| Salaried | 0.88 | 0.34 | 0.06 | 0.49 |
| Casual | 0.89 | 0.29 | 0.05 | 0.54 |
| Others | 0.88 | 0.26 | 0.06 | 0.56 |

**Table 10: Proportion of Institutional loan taken according to Household type**

| Household Type | Proportion of Institutional loan | Proportion of Institutional Housing | Proportion of Institutional Business | Proportion of Institutional Other household expenditure |
|---|---|---|---|---|
| Self-employed | 0.72 | 0.89 | 0.84 | 0.62 |

| Salaried | 0.72 | 0.93 | 0.86 | 0.67 |
|---|---|---|---|---|
| Casual | 0.58 | 0.71 | 0.92 | 0.57 |
| Others | 0.67 | 0.80 | 0.91 | 0.57 |

The tables 9 and 10 look at the proportions of loans taken according to the different employment of the household head. We find that the proportion of loan taken by each category of household type remains the same around 0.88 when it is about taking a loan or not. However when it comes to the institutional loan then self-employed and salaried household heads are better off in getting an institutional loan. This is possible because self-employed and salaried have higher incomes as compared to casuals and then their regular source of income also makes it easier.

When we look at the housing loan then salaried takes it the most followed by casuals and least are taken by self-employed, which could mean that may be the casuals take most of the informal loans. As again among the institutional housing loans also casuals are the sufferers with least proportion.
Looking at the business loans we find that self-employed households take the maximum proportion, but when we look at the institutional business loans then it's the casuals and others who take it the most. Here others are landlords, retirees, students etc. This is a bit strange because more than the self-employed, the casuals are taking more business loans.

Lastly, if we look at the other household expenditures then overall others and casuals need it the most because of their uncertain income sources they need to borrow more for their basic needs. However the trend reverses when it comes to formal other household expenditures because then self-employed and salaried takes the maximum proportion from the institutional sources. This means that casuals and others must be borrowing more from informal sources.

**Table 11: Proportion of loan according to education**

| Proportions of loans taken according to Education | | | | |
|---|---|---|---|---|
| **Education** | **Proportions of total loans** | **Proportion of total housing** | **Proportion of total Business** | **Proportion of total Other household expenditure** |
| **Illiterate** | **0.89** | **0.21** | **0.12** | **0.53** |

| Primary | 0.89 | 0.25 | 0.15 | 0.50 |
| Secondary | 0.88 | 0.31 | 0.16 | 0.45 |
| Diploma | 0.90 | 0.43 | 0.13 | 0.39 |
| Graduation & above | 0.88 | 0.49 | 0.08 | 0.38 |

Table 12: Proportion of Institutional loan according to education

| Proportions of Institutional loan taken according to Education | | | | |
|---|---|---|---|---|
| Education | Proportion of Institutional loan | Proportion of Institutional Housing | Proportion of Institutional Business | Proportion of Institutional Other household expenditure |
| Illiterate | 0.54 | 0.61 | 0.81 | 0.50 |
| Primary | 0.63 | 0.81 | 0.86 | 0.60 |
| Secondary | 0.73 | 0.9 | 0.85 | 0.68 |
| Diploma | 0.74 | 0.98 | 0.82 | 0.71 |
| Graduation & above | 0.86 | 0.97 | 0.89 | 0.75 |

The tables 11 and 12 describe the proportions of loans taken according to the education of the household head. We find that the proportion of total loans remains constant with education levels, the illiterate and graduates are taking a similar proportion of loans around 0.85. On the contrary, for taking an institutional loan, education level has a positive relationship as illiterates has a proportion of 0.54 on an average, while it is as high as 0.86 by the graduates. Coming to the different loan purposes, we find that the proportion of both total and institutional loans increases with education levels for housing loans.

However, for the other two purposes, business and other household expenditure we find that institutional loans increase with increase in education level but the overall loans falls with education. It is seen that as education level increases then business loans fall in general, which means that households prefer to take less loans, may be due to better earning capacities and even if they take a loan then they prefer it from an institutional source. Then it is seen that among household expenditure loans the maximum proportion is taken by the illiterates and the least are taken by the graduates, which reflects that education improves income and dependence on household expenditure loans decline.

## Results:

**Savings side:**

At the asset side if we look at the broad trends. Badarinza et. al (2016) finds out that in comparison to the other emerging economies and developed nations we find that Indian households tend to hold a larger proportion of their wealth in physical assets like real estate and gold instead of investing in financial assets.
On similar lines RBI (2017) the Committee on Household Finance finds out that for an average Indian household roughly eighty four percent of their wealth is kept in physical goods such as real estate, eleven percent in the form of gold and the remaining miniscule five percent in the form of financial assets. As the household approaches the top of wealth distribution then a shift is seen from the gold holdings towards real estate, without much increase in the financial assets.

Since gold and bullion deposits have a significant share in the saving portfolio of Indian households as seen in the descriptive statistic also. A recent National Household survey of gold consumption by the India Gold Policy Centre (IGPC) and People research reveals that middle income households who earn between two to ten lakh annually are the ones who invest the most in gold, even more than the rich and it accounts for about fifty six percent of the annual gold intake. It also reveals that more than seventy five percent households in rural as well as urban India owns gold in some form or another (Ghosal, 2022)

We have picked up three most relevant saving instruments that are saving accounts, use of e-wallets, gold and bullion deposits and life insurance policies. If we look at the Global Financial Inclusion (Findex) database which is drawn from 123 economies, covering 1,28,000 adults around the globe, which is close to 91 percent of the world's population. The database provides indicators about how adults manage their financial lives as well as do financial planning.

If we look at the bank account ownership at the global level we find that according to latest Findex data 2021, one-third of adults in the world, close to 1.7 billion people, were unbanked in 2017. However, we have come a long way as account ownership in adults around the world has increased from fifty one percent in 2011 to seventy six percent in 2021. On an average the account ownership has increased by eight points from sixty three percent in 2017 to seventy one percent in 2021 (Kunt et. al. 2021)

Along with account ownership, significant growth is observed in the usage of mobile money or digital payments. In developing economies, the share of adults using digital payment systems has increased from as low as thirty five percent in 2014 to fifty seven percent in 2021.In developed countries the figure is ninety five percent. Two-third of digital payment users also store money for cash management and around forty percent use it for borrowing online (Kunt et. al. 2021).

According to Global Findex Report 2021 the COVID pandemic catalyzed the growth in the usage of digital payment. In India itself around eighty million adults used digital payment mechanisms for the first time during the pandemic. In China around hundred million adults had used digital payments for the first time during the pandemic (Kunt et. al. 2021). Thus, the pandemic has given a push to digital payment mode even among the developing countries. So, while examining the financial lives in any developing country we need to evaluate the usage and patterns of digital payment gateways.

In the developing economies the households had directly leap-froged to mobile payments and digital transactions without having bank accounts, unlike in the developed economies where the individuals had opened bank accounts first and then gradually got debit and credit cards linked for mobile payments and other digital transactions (Kunt et.al. 2021). This can be considered as a step for ensuring more access to formal finance channels, however usage depends on several other factors, which would be known through the regressions.

We have seen that overall account ownership of bank accounts and overall mobile payments have increased but the growth hasn't been equal among all. There has been relatively slower progress among the women and other vulnerable groups. There is around twenty percentage point gender gap between men and women owning mobile phones in India among the other South Asian economies (Kunt et. al. 2021)

We intend to find out the factors which affect the usage of financial instruments as well as evaluate the financial choices available to the urban Indian households. So we would like to see how the three saving instruments that we have picked up respond to demographic factors like type of employment, gender, age, education, caste, religion and others. Both access and usage can differ according to the region, it depends on the district parameters like number of bank branches, average literacy and sex ratio of the overall district.

We run a simple logit model to evaluate the factors which decide whether a household uses banking service or not as well as whether e-wallets have been used or not. The dependent variable is the usage of a bank account/ e-wallets which takes the value when the household had made use of them at the time of the survey.

**Usage of bank account/ E-wallets = Household type + Household size + Gender + Age + Highest Education + Religion + Social Group + E-wallet + Income + Branches per capita + Sex Ratio + Average Literacy**

**Table 13: Whether the following have been used or not?**

|  | Bank Account | E-wallet |
|---|---|---|
| **Nature of Employment (Base: Self-employed)** | | |
| Salaried | 0.014 (0.013) | -.006 (0.012) |
| Casual | -0.016 (0.014) | **-.056* (0.018)** |
| Others | -0.009 (0.020) | 0.009 (.020) |
| **Household Size** | -0.004 (0.003) | **-.027* (.004)** |
| **Female headed Household** | 0.013 (0.015) | **-.043* (.014)** |
| **Age** | **0.001 * (0.00)** | **-.009*** |
| **Education** | | |
| Primary (Literate + primary) | **0.046 * (0.019)** | **0.033* (.005)** |
| Secondary | **0.088 * (0.019)** | **0.105* (.017)** |
| Diploma | **0.117* (0.021)** | **0.244* (.033)** |
| Graduation & above | **0.120* (0.018)** | **0.272* (.021)** |
| **Religion** | | |
| Islam | 0.020 (0.014) | 0.022 (.022) |
| Others | 0.045 * (0.019) | 0.038 (0.038) |
| **Social Group** | | |
| Schedule Caste | -0.002 (0.034) | 0.015 (0.015) |
| OBC | 0.009 (0.033) | 0.054 (.024) |
| General | 0.003 (0.035) | 0.085 (.024) |
| **E-wallets** | **-0.085 * (0.003)** | |
| **Income** | **0.037 * (0.013)** | 0.116 (.010) |
| **Branches per Capita** | 0.712 (0.479) | -2.233 (0.602) |
| **Sex Ratio** | 0.000 (0.000) | 0.00 (0.00) |
| **Average Literacy** | **-0.001 * (0.000)** | 0.00 (0.00) |

If we look at the factors which affect usage of bank accounts we find that education plays a significant and positive role in influencing households to use banks and higher education levels bring forth higher saving account usage as compared to the basic literates or with primary education. It also has a positive role in enhancing e-wallet usage as higher education leads to higher e-wallets usage.

Then income has a positive and significant effect in causing higher bank account usage but doesn't have a significant impact on e-wallets usage. Coming to the factors affecting e-wallets usage we find that casually employed households tend to use the e-wallets less than the self employed households. This could be possible because the self-employed must be accepting e-wallets payments during their business activities and the usage is higher. Then household size and female headed households are at a disadvantage because both have a negative and significant relationship with e-wallets usage.

Then we find that usage of e-wallets decline with increase in bank usage as a negative and significant relation is seen between the two. As seen in the descriptive statistic also we find a close substitutability between e-wallets and saving accounts.

In our next model we run a simple OLS regression to look at the intensity of usage (asset shares) among the bank's saving account, gold and bullion and Life insurance (LICs). We look at the amount of money kept in the savings account, amount of gold and bullion deposits and sum assured of life insurance policies as a fraction of total financial wealth. All of them will be the proportion of total financial wealth after deducting the sum assured of life insurance policies because sum assured cannot be a part of the present financial wealth of the household, it is part of the wealth when the amount is received at the maturity of the policy.

Amount in asset as proportion of total financial wealth = Household type + Household size + Gender + Age + Highest Education + Religion + Social Group + E-wallet + Branches per capita + Sex ratio + Average Literacy + Log of total financial savings + Income

**Table 14 : Asset Shares (Amount of savings as a fraction of total financial wealth)**

|  | Proportion of Saving Account | Proportion of Gold and Bullion | Proportion of LICs |
|---|---|---|---|
| **Nature of Employment: Base-Self-employed** |  |  |  |
| Salaried | **-0.049 * (0.027)** | **-.064* (0.029)** | **-.032* (0.01)** |
| Casual | **-0.068 * (0.008)** | -.026* (0.014) | **-.05* (.017)** |
| Others | 0.022 (0.014) | **.016* (0.013)** | .023 (.012) |
| **Household Size** | .001 (0.002) | 0.003 (0.003) | **.007* (.003)** |
| **Female headed Household** | -0.029 (0.007) | -0.006 (0.012) | **0.052* (.017)** |
| **Age** | **.003 * (0.00)** | **-.002*** | 0.001 (0) |
| **Education** |  |  |  |
| Primary | 0.003 (0.007) | .014 (0.013) | -.016 (.013) |
| Secondary | **0.059 * (0.008)** | .01 (0.013) | -.027 (.015) |
| Diploma | 0.019 (0.014) | .037 (0.029) | -.026 (.024) |
| Graduation & above | **0.102 * (0.01)** | **-.097* (.017)** | -.039 (.019) |
| **Religion** | | | |
| Islam | -0.09 (0.007) | -.018 (0.013) | -.015 (.011) |
| Others | 0.018 (0.018) | **-.112* (0.02)** | .005 (.01) |
| **Social Group** | | | |
| Schedule Caste | -0.001 (0.013) | .024 (0.019) | **-.105* (.027)** |
| OBC | -0.003 (0.012) | .017 (0.016) | -.017 (.024) |
| General | 0.014 (0.012) | .03 (0.017) | **-.054* (.027)** |
| **E-wallets** | **-0.067 * (0.007)** | -.014 (0.013) | -.008 (.015) |
| **Branches per Capita** | **-1.93* (0.556)** | 2.495 (0.611) | **-1.042* (.734)** |
| **Sex Ratio** | 0 (0.00) | 0 (0.00) | 0 (.00) |

| Average Literacy | 0<br>(0.00) | 0<br>(0.00) | 0<br>0 |
|---|---|---|---|
| **Total Financial Savings (log)** | **-0.032 \***<br>**(0.004)** | **-.073\***<br>**(0.007)** | **-.154\***<br>**(.009)** |
| **Income** | 0.013<br>(0.01) | **.035\***<br>**(0.018)** | **-.049\***<br>**(.017)** |

If we look at the regression results, we find that employment of the household head has a significant and negative relationship with all the three saving instruments. We find that as compared to the self-employed the salaried and casuals save less in all of them. This shows that saving capacities of the self-employed is higher than the other employment categories, which is on expected lines because when you are working for your own then scope of hiding incomes and paying less tax is common. Among the casuals, we find more uncertainty in their lives as well as their earning capacities are lower so scope of savings is lower also.

Then we find that female headed households have a positive and significant relation with life insurance policies. The female household heads in India means that the male member is missing (dead or divorced) so as head of the family they seem to be more risk averse and that's why investing in insurance policies to mitigate the financial risk from their lives as well as their dependents.

Coming to education we find that for saving in a bank account there is a positive and significant relation but the relationship becomes negative when it's about gold and bullions investment. This is so because a higher level of education encourages households to invest less in gold and more in financial investments, as evident that graduates have a negative relation with gold deposits.

Then looking at the branches per capita, we find that it has a significant relationship with all the three instruments, it has a negative relation with saving accounts and LICs but has a positive effect with gold deposits. This shows that in urban million plus agglomerations there are already enough bank branches that increasing more of it will not induce intensity of usage and moreover higher branches can lead to just opening more accounts but it cannot increase the intensity of usage. The positive relation between bank branches and gold deposit shows that despite the increase in branches the gold deposits increase which shows that it doesn't take away the charm of gold deposits or it means that households indulge in gold deposits irrespective of the availability of other saving financial instruments.

RBI (2017) the Committee on Household Finance have tried to explain the reasons for the strong liking to hold gold deposits by the Indian households. They find that the most significant reason is that gold is considered as a safe asset as it is seen as a tool for risk aversion. Then the reasons for holding gold deposits is not just because of pull factors but there are also some push factors, which are those factors which pushes households away from other financial products. They include the negative experiences faced by households while using formal financial products like social discrimination in the form of gender and caste based discrimination, long delays, high travelling time because of far off branches. Then lack of financial awareness and the misconception that financial assets are risky, they would lose their money. The study also observes that high reliance on gold deposits is not because of lack of education or financial literacy.

Income has a significant impact also, though there is a positive relationship with gold deposits as increase in income leads to increase in intensity of gold deposits. However, there is a negative relationship with intensity of LIC policies, which means that when income is higher households might prefer to take policies with lower coverage or sum assured at maturity is lower. This also reflects that with higher income households would not be so risk averse about their future financial needs.

**Loans side:**

Coming to the borrowing side, Badarinza et. al (2016) finds out that there is low reliance on institutional debt as a major proportion of household debt is taken from the non-institutional sources. They also found that there is low participation in mortgage or secured loans. RBI (2017) the Committee on Household Finance also finds that Indian households have high dependence on non-institutional debt especially from the moneylenders. The share of unsecured debt is as high as fifty six percent which is seen to trap households into long cycles of interest repayment. The committee also finds out that a household's debt follows a hump shaped curve over their life cycle, which means that they borrow more in their later lives, that is as they age, their borrowing needs increase. They keep accumulating debt till they approach retirement age, which is risky because after retirement their regular source of income shrinks.

The dependence on informal financial sources is not just in the rural areas but even for the urban areas, which means that despite availability of both institutions, informal is chosen. This could be possible because for the smaller loan amounts generally the informal is preferred due to its flexibility, convenience to the borrowers and informational advantage to the lenders over the formal channels. However, when it comes to bigger loan amounts like housing loans which require larger amounts than informal channels are inconvenient and costly due to higher rate of interest, whereas formal financial channels have the benefit of economies of scale as it deals with a larger pool of customers (Jain, 1999, Badarinza et. al. 2016-17).)

Now we look at the loan side to get an idea about the different types of loans taken by the urban households, what are the major purposes for which loans are taken, nature of interest charged and whether institutional or non-institutional loan is taken.
As we have seen in the descriptive statistics, there are four major purposes for which Indian households are borrowing that is housing loan, business loan, medical and education loan.

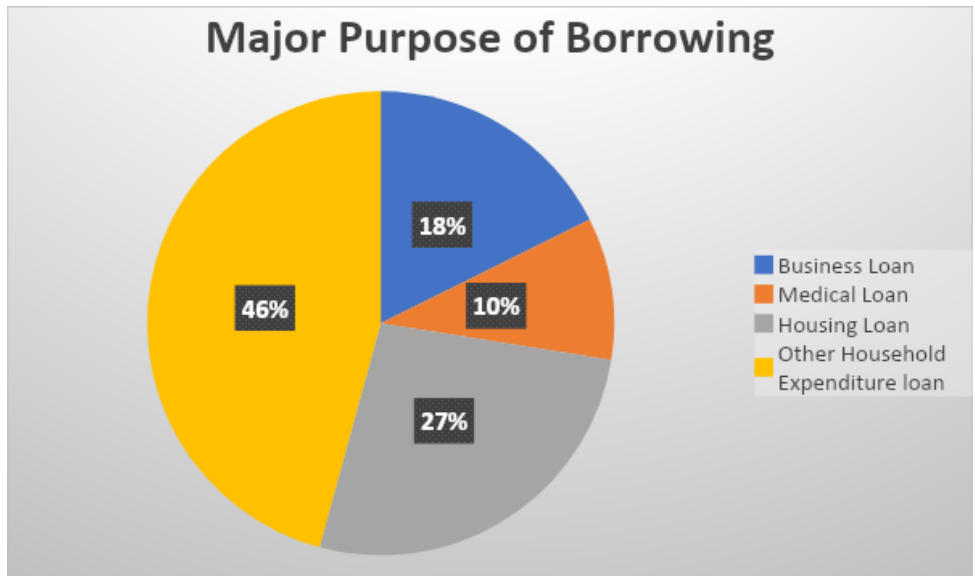
AIDIS 77th Round 2018-19

Among them it's the other household expenditure for which accounts the maximum share among the total number of loans taken by the household. The next most important reason for which loan is taken is the housing loan. The other household expenditures are largely smaller loans taken for day to day needs of the household and usually the amount borrowed is not exceeding one lakh rupees for more than seventy percent of the cases. It is usually seen that such loans are taken from the informal channels because of higher flexibility and easy terms of interest. It is for the housing or business loans that require a large sum of money and households prefer to get it from the formal financial channels with the expectation of getting lower rate of interest.

So we have picked up the prominent reasons for borrowing money by the urban Indian households and through these regressions we will get to know the leading factors which lead to dependence on institutional sources. Before anything else, first we should understand the factors which affect the household's decision of whether to take out a loan or not.

**Table 15: Has taken a loan or not**

|  | **Housing Loan** | **Business Loan** | **Other Household Expenditure loan** |
|---|---|---|---|
| **Nature of Employment** |  |  |  |
| Salaried | -0.010 (0.009) | **-0.070* (0.010)** | **0.039* (0.010)** |
| Casual | 0.005 (0.011) | **-0.067* (0.010)** | **0.060* (0.013)** |
| Others | **-0.045*** **(0.010)** | **-0.075*** **(0.011)** | **-0.022*** **(0.011)** |
| **Household Size** | 0.008 (0.002) | **0.005*** **(0.001)** | **0.006*** **(0.002)** |
| **Female headed Household** | 0.005 (0.010) | 0.006 (0.010) | 0.012 (0.013) |
| **Age** | **0.001*** **(0.00)** | -0.00 (0.000) | **0.001*** **(0.00)** |
| **Education** | | | |
| Primary | **0.003*** **(0.009)** | 0.009 (0.010) | **0.068*** **(0.012)** |
| Secondary | **0.027*** **(0.009)** | -0.010 (0.010) | **0.033*** **(0.013)** |
| Diploma | **0.083*** **(0.023)** | -0.003 (0.018) | 0.031 (0.023) |
| Graduation & above | **0.052*** **(0.011)** | -0.018 (0.011) | -0.006 (0.013) |
| **Religion** | | | |
| Islam | 0.003 (0.009) | **0.014*** **(0.005)** | -0.014 (0.013) |
| Others | 0.027 (0.019) | 0.023 (0.017) | 0.032 (0.024) |
| **Social Group** | | | |
| Schedule Caste | **0.036*** **(0.011)** | -0.003 (0.012) | **0.060*** **(0.018)** |
| OBC | **0.051*** **(0.009)** | 0.016 (0.012) | **0.071*** **(0.016)** |
| General | **0.029*** **(0.009)** | 0.003 (0.010) | 0.017 (0.016) |

We find that the employment category of the household head significantly affects the household's decision to borrow or not. As compared to the self-employed which is the base category there is a negative relationship with business loans for the salaried and casuals. This is completely on expected lines because self-employed are the ones who would need the business loans the most. On the contrary we find that salaried and casuals have a positive and significant relation with other household expenditures, reflecting more dependence for such loans as compared to the self-employed. This means that self-employed earning capacities are higher so can easily meet the household's smaller financial expenditures relatively easier than the salaried and casuals.

We find that education plays a positive and significant role in a household's decision to take a housing loan, as education level increases then the household's decision to borrow for housing increases too. When we come to the other household's expenditure then it's significant only for the lower education levels like literates or primary because later on with more education, there is less need to borrow for smaller household expenditures.

Now it's time we should understand each of these categories in detail by evaluating the factors responsible for which institutional loan is taken up, along with the intensity of usage seen through the amount of loan taken. We should look at the factors responsible for taking a housing loan from a formal source. We run a simple logit model to bring forth the factors responsible for taking a loan from an institutional agency.

We estimate the equation:

**Institutional loan = Household type + Household size + Gender + Age + Highest Education + Religion + Social Group + Type of loan + Tenure of loan + Income + Land owned + Branches per capita + Sex Ratio + Average Literacy**

**Table 16: Institutional Loan taken or not**

| Nature of Employment | Housing Loan | Business Loan | Other Household Expenditure loan |
|---|---|---|---|
| Salaried | 0.025 (0.039) | 0.035 (0.043) | 0.031 (0.040) |
| Casual | -0.028 (0.051) | **0.101*** **(0.032)** | -0.007 (0.040) |
| Others | -0.001 (0.089) | **0.118*** **(0.031)** | -0.003 (0.054) |
| **Household Size** | 0.001<br>(0.012) | **0.015*<br>(0.009)** | **0.027*<br>(0.010)** |
| **Female headed Household** | -0.051<br>(0.052) | **0.076*<br>(0.036)** | 0.024<br>(0.040) |
| **Age** | -0.000<br>(0.003) | 0.001<br>(0.003) | 0.001<br>(0.001) |
| **Education** | | | |
| Primary | -0.009<br>(0.049) | 0.075<br>(0.058) | 0.069<br>(0.060) |
| Secondary | -0.012<br>(0.052) | 0.072<br>(0.069) | **0.133*<br>(0.063)** |
| Diploma | -0.043<br>(0.080) | 0.102<br>(0.083) | 0.151<br>(0.099) |
| Graduation & above | 0.049<br>(0.049) | **0.143*<br>(0.063)** | **0.132*<br>(0.068)** |
| **Religion** | | | |
| Islam | 0.124<br>(0.093) | 0.068<br>(0.043) | **0.178*<br>(0.044)** |
| Others | 0.182<br>(0.112) | **0.114*<br>(0.053)** | **0.269*<br>(0.068)** |
| **Social Group** | | | |
| Schedule Caste | 0.070<br>(0.056) | **-0.130*<br>(0.061)** | 0.082<br>(0.077) |
| OBC | -0.010<br>(0.052) | **-0.078*<br>(0.028)** | 0.096<br>(0.073) |
| General | 0.059<br>(0.054) | **-0.072*<br>(0.033)** | 0.103<br>(0.078) |

| Type of loan | **0.159\***<br>**(0.059)** | **0.066\***<br>**(0.043)** | **0.108\***<br>**(0.040)** |
|---|---|---|---|
| Tenure of loan | 0.012<br>(0.092) | 0.034<br>(0.063) | 0.009<br>(0.040) |
| Income | **0.155\***<br>**(0.039)** | **0.079\***<br>**(0.043)** | **0.099\***<br>**(0.030)** |
| Land owned | **0.101\***<br>**(0.049)** | 0.001<br>(0.023) | 0.006<br>(0.033) |
| Branches per Capita | 0.000<br>(0.001) | **5.927\***<br>**(0.283)** | -0.021<br>(1.590) |
| Sex Ratio | -0.000<br>(0.00) | **-0.001\***<br>**(0.00)** | 0.000<br>(0.000) |
| Average Literacy | **0.004\***<br>**(0.001)** | -0.001<br>(0.00) | **0.005\***<br>**(0.001)** |

We find that the type of employment has a positive and significant effect only for the business loans from an institutional source. The effect is positive only for the casuals and other categories as compared to the self-employed which is the base category. This is not on expected lines as it means that casuals and those in the other categories like retirees or landlords can get an institutional business loan more easily as compared to the self-employed. This could mean that either the casuals or others must be doing a part time business for which they seek funding in the form of a loan and they must be having some collateral which can be backed up for getting an institutional loan.

Then we find that the female headed households have a positive and significant relation with institutional business loan which means that when women becomes the head of the family by choice or compulsion (divorced or male member died) then she take up the role and would become self-employed in the form of opening a grocery store or running a beauty salon as seen among the low-income households under considerations. Then in all such situations women can easily get a formal loan for the same. This is generally not on expected lines because usually gender divide is seen in the literature.

There has been considerable evidence which shows that gender plays a significant role in financial development thereby leading to economic development (Duflo, 2012). Ghosh and Vinod (2017) finds that gender divide is present in both financial access as well as usage among the women in India using All India Debt and Investment survey 70[th] round 2013-14 and it shows that female headed households are eight percent less likely to have access to finance and are more likely to resort to informal finance by six percentage points after incorporating all household and state level indicators. The study further shows that education and income play a significant role in determining financial access while social factors play a significant role in usage of financial services by the female headed households.

Then household size also has a positive and significant impact on getting an institutional business and other household expenditure loan which is true because more family members lead to more expenditures as well as more earning members depending on the number of dependents in the household. That's why it increases business loans if there are more adults who need money to be self-employed and overall more members means more financial requirements to meet the household expenditures. It's a good thing that institutional loans can be taken easily for both these purposes among the urban agglomerations.

Then we find that type of loan has a positive and significant effect for all the three loan purposes. This type of loan refers to those loans which are exceeding one lakh rupees. This means that as loan amount increases then dependence on institutional finance increases irrespective of the loan category. We also find income has a positive and significant impact for all the categories of loans, this is being reinforced here as descriptive statistics have shown the same effect.

We find that land ownership has a positive effect on the housing loan alone which is true because households need to own a land first to be able to take a housing loan for the same. However, this also shows that land ownership which can be a collateral is not a prerequisite for getting a formal loan as it is not significant for other loan purposes. Lastly, we find that branches per capita have a positive and significant relation with institutional business loans. This shows that as bank branches are higher then probability of getting an institutional loan gets increased by 5.92 units.

Next we estimate the amount of institutional loan taken as a fraction of income is regressed over the possible factors. This would help us know what are the factors which determine the intensity of loan being provided by the formal financial institution. The amount of loan disbursed by the institutional sources as a fraction of household income will depend on the demographic profile as well as some broad level district factors.

We estimate the equation:

**Institutional loan amount as a fraction of income = Household type + Household size + Gender + Age + Highest Education + Religion + Social Group + Type of loan + Tenure of loan + Income + Land owned + Branches per capita + Sex Ratio + Average Literacy**

**Table 17: Intensity of Loan (Amount of institutional loan as a fraction of income/annualcons)**

|  | **Housing Loan** | **Business Loan** | **Other Household Expenditure loan** |
|---|---|---|---|
|  |  |  |  |

| Nature of Employment | | | |
|---|---|---|---|
| Salaried | -0.267 (0.893) | -.643 (0.523) | .059 (.118) |
| Casual | -0.848 (1.026) | -.261 (0.323) | .266 (.213) |
| Others | **-3.354* (0.242)** | **-1.386 * (0.751)** | -.084 (.242) |
| **Household Size** | 0.267 (0.242) | .093 (0.155) | **-.022* (.031)** |
| **Female headed Household** | -0.521 (0.742) | 1.682 (0.155) | -.245 (.132) |
| **Age** | -0.008 (0.03) | -.026 (0.23) | 0.009 (.006) |
| **Education** | | | |
| Primary | 1.033 (0.978) | -.122 (0.325) | -.095 (.101) |
| Secondary | 0.969 (1.02) | **1.25 * (0.643)** | **.013* (.158)** |
| Diploma | **3.55* (0.978)** | -.579 (0.623) | .263 (.255) |
| Graduation & above | **3.829*** **(1.155)** | **1.543 *** **(0.698)** | **.922*** **(.922)** |
| **Religion** | | | |
| Islam | 1.088 (0.932) | .058 (0.253) | **.082*** **(.183)** |
| Others | 0.104 (1.225) | .787 (0.523) | **-.163*** **(.195)** |
| **Social Group** | | | |
| Schedule Caste | 1.040 (1.16) | -.811 (0.523) | -.119 (.183) |
| OBC | 1.69 (1.15) | -.361 (0.321) | .079 (.206) |
| General | **1.921*** **(0.883)** | -.823 (0.717) | .131 (.203) |
| **Type of loan** | **4.674*** **(0.693)** | **2.302 *** **(0.282)** | **1.108*** **(.229)** |
| **Tenure of loan** | 1.62 (0.982) | 0.747 (0.46) | 0 (.00) |
| **Income** | -0.983 (0.039) | **-1.342 *** **(0.739)** | **-0.983*** **(.58)** |
| **Land owned** | 0.101 (0.049) | **1.145 *** **(0.675)** | 0.193 (.85) |
| **Branches per Capita** | 0.002 (0.001) | .002 (0.003) | 14.212 11.15 |
| **Sex Ratio** | -0.005 (0.00) | .004 (0.003) | -.001 (.008) |
| **Average Literacy** | 0.034 (0.001) | -.028 (0.002) | **.009\* (.005)** |

We find that the type of employment of the household head is significant and negatively related with the amount of borrowing done as a proportion of income with the others among them. This shows that in comparison to the self-employed the others would be able to borrow less for obvious reasons that they consist of retirees, students and landlords who do not need to borrow that much.

Next, we find that education has a positive and significant role in influencing higher intensity of institutional borrowing for all the three categories of loan. We find that for graduates and above education level, there is more dependence on institutional borrowing. This reinforces the findings by Badarinza et. al (2016) which reiterates that education plays an important role in shifting households towards low cost institutional debt sources and helps in reducing the overall high dependence on non-institutional debt in general.

Again we find that type of loan is positive and significantly related with intensity of borrowing for all the three categories. This shows that as the loan amount is higher it leads to an increase in the amount of institutional borrowing. This again reinforces that for higher loan amounts households prefer to borrow institutionally.
Next we have seen earlier that income has a positive effect in influencing institutional borrowing however when it comes to intensity of borrowing then except for housing loans which involves very large amounts for other loan categories we find that as income of the household increases then borrowing intensity reduces. This is possible because higher income levels leads to higher earning capacities and reduces the amount of borrowing specially for purposes like household expenditures.

Lastly, we find that the intensity of borrowing is higher for households who have land ownership, that is those households who have the required collateral then the amount of business loan provided is higher. So collateral is useful in increasing the intensity of borrowing.

## **Conclusion:**

All in all we find that the income, education, type of employment and branches per capita are the main significant factors which affect the financial behaviour of the low-income households living in the million plus urban agglomerations. These results provide us a broad picture of how the financial behaviour of low-income households is affected in the top urban regions in the Indian economy.

We find that there is significant correlation with household characteristics. Firstly, education plays a significant and positive role in increasing household asset allocation towards financial assets in the form of bank account savings and e-wallets over physical assets that are gold and bullion deposits as a fraction of total savings. Education has a positive and significant role in increasing institutional borrowing for all the three loan purposes that is for housing, business and household expenditures and also positively influences the intensity of usage also. The more educated households indulge in higher borrowing amount

Secondly, we find that the type of employment of the household head significantly affects the asset allocation as well significantly changes according to the purpose and source of borrowing by the household. We find that the self-employed households have a higher usage as well as intensity of usage for financial assets rather than physical assets. It is observed that this type of employment has a positive and significant impact on usage of e-wallets and saving accounts and it is insignificant for the gold deposits as a proportion of total savings.

If we break up the type of employment we find that usage of financial assets as well as its intensity of usage is much higher for self-employed households as compared to its casual and salaried household heads. This trend is not on expected lines because it is perceived that those who are salaried must be working in either public or private firms and they are the ones who might be getting online salaries or wages in their bank account and their chances for using digital payment mechanisms would be higher. However, we find that the self-employed are using bank accounts more actively and also make use of digital payments through e-wallets more than their salaried and casual counterparts.

We find that self-employed households are at a disadvantage when it comes to getting a formal loan as compared to its salaried and casual counterparts. This is actually in reverse of what we get from the asset side, that the self-employed household heads were using bank accounts more intensively as well as using digital payments. It could mean that when it comes to borrowing then self-employed household heads might be getting several informal borrowing options and they prefer them over formal borrowing.
In case of intensity of usage, we find that intensity is much higher in case of self-employed people as compared to their counterparts that are salaried and casual labourers.

Thirdly, we find the presence of gender divide in both the cases of savings as well as in the case of borrowings. The female headed households are at a disadvantage when it comes to owning bank accounts to owning mobile phones as well as in getting formal loans. However, the gender effect works in favour of females only at the time of investment in gold and bullion. When it comes to getting loans then also women headed households are at a disadvantage than their male counterparts.

Fourthly, we find that social grouping, that is the caste of the household and religion of the household has a significant effect in case of credit disbursement. It is seen that due to their own religious beliefs the households belonging to Islamic religion are at a disadvantage when it comes to opting for formal financial institutions for borrowings. This is because of their own religious belief that is they restrict themselves from borrowing from anywhere be it formal or informal channels. Same holds true for households belonging to scheduled tribes as they also don't also indulge in borrowing due to their customs and beliefs.

Fifthly, We find income has a positive and significant impact in enhancing usage of financial assets like bank accounts, gold and bullion deposits and life insurance policies. However, when it comes to loans then we find income has a positive relationship in influencing institutional usage but when we look at the intensity of usage then the amount of loan borrowed has a negative relationship with income.

So these household characteristics like education level, social characteristics like caste, type of employment and religion, economic characteristics like income levels significantly affect the financial behaviour of Indian households. Thus to understand these relations more closely we have chosen to study an Urban metropolis Delhi, where we specifically study three unauthorised colonies where the low income and education levels along with informality in employment is there. Being in an urban metropolis we assume that access to formal or institutional finance should not be difficult for those households, so it will be interesting to study how their economic and social characteristics impact their usage of financial instruments as well as their overall financial behaviour.